\documentclass[twocolumn]{aastex61}

\accepted{for publication in ApJ }
\shorttitle{ECM emission from low mass stars}
\shortauthors{Llama et al.}
\begin{document}

\title{Simulating radio emission from Low Mass Stars}

\correspondingauthor{Joe Llama}
\email{joe.llama@lowell.edu}

% \author{Joe Llama\altaffilmark{1}}
% \author{Moira M. Jardine\altaffilmark{2}}
% \author{Kenneth Wood\altaffilmark{2}}
% \author{Gregg Hallinan\altaffilmark{3}}
% \author{Julien Morin\altaffilmark{4}}

% \altaffiltext{1}{Lowell Observatory, 1400 W. Mars Hill Rd. Flagstaff. Arizona. 86001. USA.}
% \altaffiltext{2}{SUPA, School of Physics \& Astronomy, University of St Andrews, North Haugh. St Andrews. Fife. KY16 9SS. UK.}
% \altaffiltext{3}{California Institute of Technology, 1200 East California Boulevard, Pasadena, California 91125, USA}
% \altaffiltext{4}{LUPM, Universit\'e de Montpellier, CNRS, place E. Bataillon, F-34095, Montpellier, France}
\author[0000-0003-4450-0368]{Joe Llama}
\affiliation{Lowell Observatory, 1400 W. Mars Hill Rd. Flagstaff. Arizona. 86001. USA.}
% \nocollaboration
\author[0000-0002-1466-5236]{Moira M. Jardine}
\affiliation{SUPA, School of Physics \& Astronomy, University of St Andrews, North Haugh. St Andrews. Fife. KY16 9SS. UK.}
\author{Kenneth Wood}
\affiliation{SUPA, School of Physics \& Astronomy, University of St Andrews, North Haugh. St Andrews. Fife. KY16 9SS. UK.}
% \nocollaboration
\author{Gregg Hallinan}
\affiliation{California Institute of Technology, 1200 East California Boulevard, Pasadena, California 91125, USA}
% \nocollaboration
\author[0000-0002-4996-6901]{Julien Morin}
\affiliation{LUPM, Universit\'e de Montpellier, CNRS, place E. Bataillon, F-34095, Montpellier, France}
% \nocollaboration



\begin{abstract}
Understanding the origins of stellar radio emission can provide invaluable insight into the strength and geometry of stellar magnetic fields and the resultant space weather environment experienced by exoplanets. Here, we present the first model capable of predicting radio emission through the electron cyclotron maser instability using observed stellar magnetic maps of low mass stars. We determine the structure of the coronal magnetic field and plasma using spectropolarimetric observations of the surface magnetic fields and the X-ray emission measure. We then model the emission of photons from the locations within the corona that satisfy the conditions for electron cyclotron maser emission. Our model predicts the frequency, and intensity of radio photons from within the stellar corona.

We have benchmarked our model against the low mass star V374 Peg. This star has both radio observations from the Very Large Array and a nearly simultaneous magnetic map. Using our model we are able to fit the radio observations of V374 Peg, providing additional evidence that the radio emission observed from low mass stars may originate from the electron cyclotron maser instability. Our model can now be extended to all stars with observed magnetic maps to predict the expected frequency and variability of stellar radio emission in an effort to understand and guide future radio observations of low mass stars.
\end{abstract}

\keywords{stars: individual (V374 Peg) -- stars: low-mass -- stars: activity -- stars: magnetic field}



\section{Introduction} \label{sec:intro}
One of the primary drivers in determining the space weather environment of a close-in exoplanet is the stellar magnetic field and wind. For planets orbiting M stars this is of critical importance when considering their potential for habitability. Due to their lower mass (0.1--0.6 M$_\odot$), these stars are less luminous than solar type stars, which in turn means the habitable zone is located much nearer to the star at a distance of $\sim$0.1--0.4 au \citep{Kopparapu2013}. This distance makes it easier for us to detect planets orbiting within the habitable zone; however, these planets may be subjected to more frequent and intense space weather conditions than any of the planets in our solar system (e.g., \citealt{khodachenko2007,vidotto2013,cohen2014,see2014,cohen2015,garraffo2017}).

In the solar system, the auroral regions of magnetized planets emit coherent, bright, polarized,  low-frequency radio emission through the electron cyclotron maser instability (ECM; \citealt{farrell1999,zarka1998,ergun2000,treumann2006,hallinan2013}) where electrons are accelerated along the planet's magnetic field lines. The power of this emission has been shown to scale directly with the incident power of the solar wind that interacts with the magnetospheric cross-section of the planet. This relation, known as the ``Radio Bode's law'' spans many orders of magnitude in the solar system planets \citep{farrell1999,zarka2001}.

With over 3000 exoplanets discovered to-date\footnote{\url{http://www.exoplanets.org}} there has been considerable effort to detect radio emission from these planets. A successful detection would allow us to directly measure the magnetic field strength of the planet which so far has only been done through indirect measurements of star-planet interactions (e.g., \citealt{shkolnik2005,shkolnik2008,vidotto2010,llama2011,haswell2012,gurdemir2012}). Exoplanetary magnetic fields provide insight into the internal structure and composition of the planet and potentially play a crucial role in habitability, shielding the planet from energetic particles from the stellar wind and from cosmic rays. Radio emission also offers an alternative method for directly detecting exoplanets \citep{farrell1999}.

Extrapolations of Bode's law to exoplanets have suggested that due to their small orbital separations, hot Jupiters should emit radio emission at levels orders of magnitude greater than Jupiter in our solar system \citep{lazio2004}. The promise of bright radio emission from exoplanets has prompted many searches; however, these have mostly yielded null detections \citep{ryabov2004,lazio2007,hallinan2013}. A search for the secondary eclipse of the transiting planet HD 189733b by \citet{smith2009} provided an upper limit at 307-347 MHz, while observations of the HAT-P-11 system by \citet{lecaval2013} found a tentative detection of 150 MHz emission from HAT-P-11b. An extensive 150 MHz survey by \citet{sirothia2014} found null detections from the 61 Vir system, which was predicted to be radio bright and also the 55 Cnc system. At 1.4 GHz, \citet{sirothia2014} made a tentative detection from the planet harboring pulsar PSR B1620-26, WASP-77 A b, and HD 43197b. A recent 2--4 and 4--8 GHz search for radio emission from $\epsilon$ Eridani b was carried out by \citet{bastian2017}; however, they could not definitively determine whether the source of the observed radio emission was from the planet.

The lack of a detection of radio emission from exoplanets is likely due to these surveys being less sensitive to the frequencies predicted from Bode's law \citep{farrell1999,bastian2000,lanza2009,lazio2009,jardine2008,vidotto2012,lazio2016}. Since the radio flux scales directly with the power of the incident stellar wind, targeting young systems which host dense, strong stellar winds may offer an exciting opportunity to make a definitive detection of radio emission from exoplanets. Indeed, there are now a number of planets known around young stars, including CI Tau b \citep{johnskrull2016}, V830 Tau b \citep{donati2016}, K2-33 b \citep{mann2016,david2016}, and TAP 26 b \citep{yu2017}. \citet{vidotto2017} carried out a theoretical study to predict the radio emission from V830 Tau b, a $\sim2$ Myr old hot Jupiter orbiting a pre-main sequence star. By simulating the stellar wind of V830 Tau using three-dimensional MHD models coupled with magnetic imaging of the host star these authors estimate the radio flux density from V830 Tau b to be 6 -- 24 mJy.

Low mass stars with spectral type later than $\sim$M4 are fully convective, meaning they lack a radiative core and a tachocline (the interface layer between the radiative core and the convective outer envelope). At even lower masses, the ultracool dwarfs (UCDs) with a spectral type $\ge$ M7 that populate the very end of the main-sequence  represent a change in magnetic activity. These objects are of particular interest because they span the boundary between stars and hot Jupiters. X-ray observations have shown that the bolometric levels of X-ray emission, $L_X/L_{\rm bol}$, decrease by two orders of magnitude, suggesting they do not host a magnetically heated corona \citep{mohanty2002,stelzer2006,reiners2008,berger2010}. Despite the lack of X-ray emission, radio observations have revealed strong emission for UCDs spanning late M through to T dwarfs, suggesting that these stars are capable of maintaining strong magnetic fields \citep{hallinan2008,berger2010,mclean2011,williams2013,williams2017,route2016}.

Radio observations of LSRJ1835+3259, an M8.5 star with a 2.84 h rotation period found pulsed radio emission that also phased with their simultaneous optical Balmer observations \citep{hallinan2015}. From the frequency of this emission \citet{hallinan2015} were able to determine that the star hosts a magnetic field between $B\sim1,550 - 2,850$ Gauss. Both the pulses and also the background emission from UCDs have been attributed to ECM emission \citep{hallinan2006,hallinan2008}. This instability is also believed to power the ``stellar auroral emission'' seen in the massive star CU Vir \citep{trigilio2004,leto2006,leto2016}.

Our understanding of how the dynamo magnetic field in fully convective, low mass stars is generated is far from complete; however, magnetic imaging of bright, rapidly rotating stars through Zeeman Doppler Imaging (ZDI; \citealt{semel1989,donati1997,donati2006}) is allowing us to study the topology and evolution of stellar magnetic fields for a wide range of stars the pre- and main-sequence through surveys such as BCool (solar type stars; \citealt{marsden2014}), MAPP (classical T Tauri stars; \citealt{donati2012}), MiMeS (massive stars; \citealt{wade2016}), MaTYSSE (young planet hosting stars; \citealt{donati2014}), and BinaMIcS (short period binary stars; \citealt{alecian2015,alecian2016}). To map the full magnetic topology of a star, polarized spectra are collected during at least one rotation of the star. The technique is therefore most suitable for stars with rapid rotation periods. ZDI observations of low mass stars have revealed that M0--M4 stars have weak large-scale magnetic fields while stars later than M4 host large-scale fields that may be either strong and axisymmetric or weak and complex \citep{morin2008}.

% M stars are the most abundant in the galaxy and the current favored targets for finding a potentially habitable exoplanet. Due to their lower mass (0.1--0.6 M$_\odot$), these stars have much longer main-sequence lifetimes than solar type stars. Their lower mass also means they are less luminous, which in turn means the habitable zone is located much nearer to the star at a distance of $\sim$0.1--0.4 au \citep{Kopparapu2013}. This distance makes it easier for us to detect planets orbiting within the habitable zone; however, these planets will be subjected to much stronger space weather conditions than any of the planets in our Solar System.

% At such short orbital separations close-in exoplanets are likely being exposed to strong levels of stellar X-ray and UV flux, which will heavily irradiate the planet, heating the upper atmosphere, which in turn alters the chemical composition causing hydrodynamical blow-off \citep{vidalmadjar2003,vidalmadjar2004,benjaffel2010,lecaval2010,lecaval2012,bourrier2013,ehrenreich2012,kulow2014,ehrenreich2015}.

One low mass star that sits right on the boundary of being fully convective is V374 Peg. This low mass ($M_\star=0.28M_\odot,\,r_\star=0.34r_\odot$) star is located in the nearby stellar neighborhood ($d=8.93$ pc; \citealt{vanleeuwen2007}) and is rapidly rotating ($P_{\rm rot}=0.44$ d; \citealt{morin2008v374}). V374 Peg has been observed over many years, and shows signs of frequent flaring and magnetic activity (e.g., \citealt{batyrshinova2001,korhonen2010,vida2016}). Given its proximity and rapid rotation, V374 Peg is an ideal candidate for magnetic imaging through ZDI. \deleted{ZDI is a tomographic technique that exploits the polarization of magnetically sensitive lines in the photosphere to measure the global topology of the large scale stellar magnetic field.}

Magnetic maps for V374 Peg were obtained on two epochs, first in 2005 August and September \citep{donati2006} and again a year later in 2006 August \citep{morin2008v374}. The magnetic topology of V374 Peg was found to be predominantly dipolar with a peak field strength of $B_0=1,660$ G. \citet{vidotto2011} used the ZDI maps as input into a 3D MHD model to compute the stellar wind properties of V374 Peg, finding that the star has a fast, dense wind with a ram pressure five times larger than the solar wind. V374 Peg is also radio bright, exhibiting a rotationally modulated but smoothly varying component of emission, coupled with pulsed radio bursts that phase with the rotation period of the star \citep{hallinan2009}.

In this paper we present the first model that couples stellar magnetic maps (observed and reconstructed using ZDI) with a model to predict the amplitude, variability, and frequency of ECM emission. In Section \ref{sec:model} we describe our model for simulating radio emission through ECM, including an overview of the potential field source surface extrapolation that enables us to compute the properties of the stellar corona from a ZDI map. In Section \ref{sec:results} we present the results of applying the model to a) a simple inclined dipole magnetic field and b) to the magnetic map of the M dwarf V374 Peg. In Section \ref{sec:v374_obs} we compare the predicted ECM radio light curve for V374 Peg with near simultaneous data obtained from the Very Large Array (VLA) and show that our model is capable of reproducing both the variability and amplitude of the observations.

\section{The Model}\label{sec:model}
\subsection{Stellar magnetic field and wind}\label{sec:pfss}
ZDI observations provide a topological map of the surface distribution of the large-scale stellar magnetic field. From these maps we can determine the structure of the stellar corona by applying a potential field source surface model (PFSS; \citealt{altschuler1969,jardine2002}). \replaced{This approach assumes the magnetic field to be in a potential state, i.e., $\nabla\times \textbf{B} = 0$ so that the field can be expressed as a scalar potential, $\textbf{B}=-\nabla \varphi$. Upon substitution into Gauss' law, $\nabla\cdot\textbf{B}=0$, the scalar potential must satisfy Laplace's equation, $\nabla^2\varphi=0$. The radial, meridional, and azimuthal components of the magnetic field can then be expressed as a sum of spherical harmonics in terms of the associated Legendre polynomials $P_{lm}$, where $l$ indicates the spherical harmonic degree, and $m$ indicates the order as,
% \begin{eqnarray}
%     B_r = -\sum_{l=1}^N\sum_{m=1}^{l}&&\left[la_{lm}r^{l-1}-(l+1)b_{lm}r^{-(l+2)}\right] P_{lm}(\cos\theta)e^{im\phi}, \\
%     B_\theta = -\sum_{l=1}^N\sum_{m=1}^{l}&&\left[a_{lm}r^{l-1}-b_{lm}r^{-(l+2)}\right] \frac{{\rm d}}{{\rm d}\theta}P_{lm}(\cos\theta)e^{im\phi}, \\
% B_\phi = -\sum_{l=1}^N\sum_{m=1}^{l}&&\left[a_{lm}r^{l-1}-b_{lm}r^{-(l+2)}\right] P_{lm}(\cos\theta)\frac{im}{\sin\theta}e^{im\phi},
% %     B_r = -\sum_{l=1}^N\sum_{m=1}^{l}&&\left[la_{lm}r^{l-1}-(l+1)b_{lm}r^{-(l+2)}\right]\times\\
% %     \nonumber&&  P_{lm}(\cos\theta)e^{im\phi}, \\
% %     B_\theta = -\sum_{l=1}^N\sum_{m=1}^{l}&&\left[a_{lm}r^{l-1}-b_{lm}r^{-(l+2)}\right]\times\\
% %     \nonumber && \frac{{\rm d}}{{\rm d}\theta}P_{lm}(\cos\theta)e^{im\phi}, \\
% % B_\phi = -\sum_{l=1}^N\sum_{m=1}^{l}&&\left[a_{lm}r^{l-1}-b_{lm}r^{-(l+2)}\right]\times\\
% %  \nonumber && P_{lm}(\cos\theta)\frac{im}{\sin\theta}e^{im\phi},
% \end{eqnarray}
where $a_{lm}$ and $b_{lm}$ are the amplitudes of each spherical harmonic component. The values of $a_{lm}$ and $b_{lm}$ are derived by prescribing two boundary conditions on the model, one at the stellar surface, $r=r_\star$, and one at the source surface, $r = r_{\rm ss}$.}{This approach assumes the magnetic field to be in a potential state and requires the prescription of two boundary conditions, one at the stellar surface, $r=r_\star$, and one at the source surface, $r = r_{\rm ss}$.} The boundary condition at $r=r_\star$ is set to the radial component of the magnetic field obtained through ZDI. At $r=r_{\rm ss}$, the boundary condition that the magnetic field becomes purely radial, i.e., $B_\theta=B_\phi=0$ is imposed. This condition is analogous to imposing the maximum extent of the closed corona, and beyond the source surface the field is entirely open, carrying the stellar wind. While it is not possible to observe the extent of the closed corona  for stars other than the Sun, dynamo simulations have shown that it likely varies with the fundamental parameters of the star (e.g., \citealt{reville2015}). Here we adopt the solar value of $r_{\rm ss}=2.5r_\star$; however, we also ran simulations with $r_{\rm ss}=5r_\star$ with negligible differences between those presented here.

\subsection{Modeling the coronal density structure}\label{sec:xray}
From the magnetic field extrapolation we next determine the density structure of the stellar corona. We assume that the coronal plasma is in isothermal, hydrostatic balance, such that the pressure on each closed field line is given by:
\begin{equation}
p = p_0\exp\left[(m/kT)\int g_s\, {\rm d}s\right],
\end{equation}
where $p_0$ is the plasma pressure at the base of the field line (which we set to $p_0=\kappa B_0^2$), where $\kappa$ is a scaling parameter, $B_0$ is the field strength at the base of the field line and $g_s = (\textbf{g}\cdot\textbf{B})/|\textbf{B}| $ is the component of the effective gravity along the field line. If along any field line the plasma pressure is greater than the magnetic pressure, we assume that the field line would have been forced open by the plasma pressure and we set the pressure to zero. This is also the value used for open field lines. Once the pressure is known the density can be determined by assuming an ideal gas. We then carry out a Monte-Carlo radiation transfer simulation to produce a 3D model of the X-ray corona \citep{wood1999}. We assume the \replaced{emission is optically thick and so }{emissivity} scales \replaced{as the square of the }{directly with the}  local coronal density.

\subsection{Modeling Radio Emission}\label{sec:radio}
In this work we are interested in simulating radio emission through the ECM instability. \added{Using the formalism of \citet{treumann2006}, the most efficient condition for ECM emission is that the local electron plasma frequency,
\begin{equation}
    \omega_p=\frac{e}{2\pi}\sqrt{\frac{n_e}{m_e\epsilon_0}}\approx 9\sqrt{n_e}\,\,\textrm{kHz},
\end{equation}
should be less than the electron-cyclotron frequency,
\begin{equation}
    \Omega_e = \frac{eB}{2\pi m_e} \approx 28\,B\,\,\textrm{kHz},
\end{equation}
where $n_e$ is the electron plasma density, and $B$ is the local magnetic field strength. The ECM mechanism is most efficient for $\omega_p^2/\Omega_e^2 \ll 1$; however, it will also work (just less efficiently) under the condition $\omega_p^2/\Omega_e^2 <1$. For completeness we therefore allow ECM emission from all sites in the coronal volume where
\begin{equation}
    \sqrt{\frac{n_eq^2}{m_e\epsilon_0}} < \frac{qB}{m_e}.
    \label{eqn:ecm1}
\end{equation}}
 %This condition is typically expressed in terms of the Alfv\'en speed,
%\begin{equation}
%    v_A = \frac{B}{\sqrt{\mu_0n_im_i}},
%    \label{eqn:va}
%\end{equation}
%where $n_i$ and $m_i$ are the ion density and mass respectively.
% Assuming $n_i=n_e$, the ratio $\omega_p/\Omega_e$ can then be expressed as:
% \begin{equation}
%     \frac{\omega_p}{\Omega_e} = \sqrt{\frac{m_e}{m_i}}\frac{c}{v_A}.
%     \label{eqn:ecm}
% \end{equation}
%so that, for example, the condition for emission from a fully-ionised Hydrogen plasma would be $v_A > 7000$kms$^{-1}$.
In terms of the local variables determined by our coronal model, this can most usefully be written as
\begin{equation}
     \left[\frac{n_e }{10^{14} \rm{m}^{-3}} \right]< \left( \frac{28}{9} \right)^2  \left[\frac{B}{100 \rm{G}} \right]^2,
     \label{eqn:condition}
\end{equation}
Regions of low density and high field strength are the most likely to emit. Locations in the coronal volume where Equation (\ref{eqn:condition}) is satisfied emit photons at the local gryo frequency:
\begin{equation}
    \nu = \frac{qB}{2\pi m}.
    \label{eqn:gyrofreq}
\end{equation}
 For electrons, Equation (\ref{eqn:gyrofreq}) can be expressed as $\nu_{\rm MHz} \approx 2.8 \times B_{\rm Gauss}$. We assume that photons are emitted into a hollow cone distribution, where the thickness of the cone is $1^\circ$ and the opening angle is $90^\circ$ \citep{melrose1982}. \replaced{The intensity, $I(\textbf{r},\nu)$ is therefore given by}{The number of electrons that can emit towards the observer at rotation phase $\phi$, and frequency $\nu$ is given by }
 \begin{equation}
    N(\phi, \nu) = \sum_{i}\exp\left(\frac{-\Delta\theta_i^2}{2\sigma^2}\right)n_i(\nu)\,dV_i,
\end{equation}
%  \begin{equation}
%     \epsilon(\underline{r},\nu) = \sum_{i}\exp\left(\frac{-\Delta\theta_i^2}{2\sigma^2}\right)n_i(\textbf{r},\nu)\,dV_i,
%     \label{eqn:rI}
% \end{equation}

where $\Delta\theta_i$ is the angle between the magnetic field and the plane of the sky $\sigma$ is the thickness of the cone, $n_i(\nu)$ is the number of electrons in grid cell $i$ with frequency $\nu$ that can emit ECM photons, and $dV_i$ is the volume of the grid cell. We assume the star is optically thick and set all grid cells that are behind the star to zero. The polarization of the radio emission is determined by the sign of the local radial magnetic field.
\section{Results}\label{sec:results}

\subsection{Simple case: Dipolar Magnetic Field}
\begin{figure*}
\centering
    \includegraphics[width=1\textwidth]{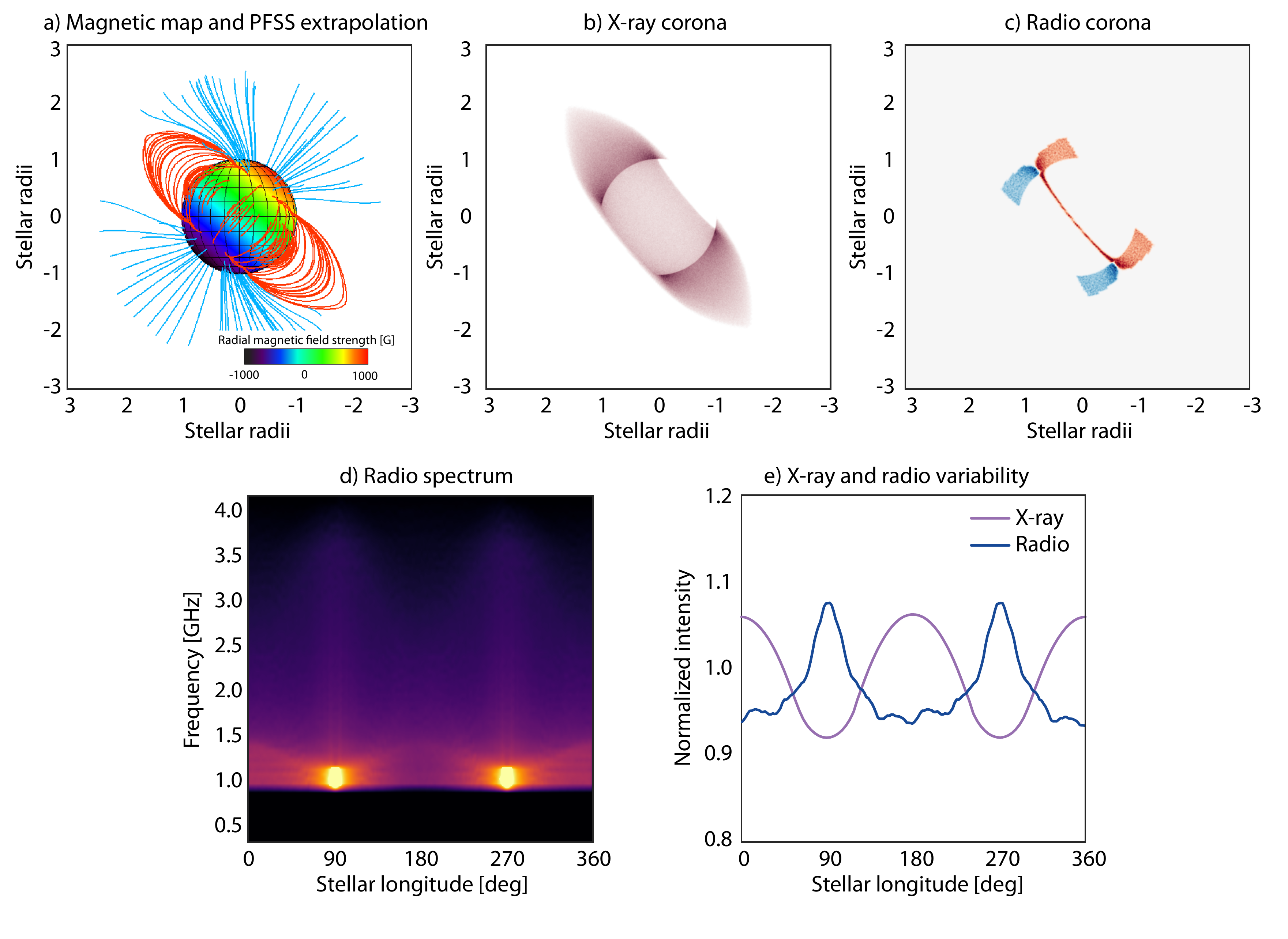}
    \caption{a) Simulated magnetic map of an inclined dipole ($B_0=1,000$ G, $\beta=40^\circ$) and the PFSS model (Section \ref{sec:pfss}). b) X-ray coronal density structure (Section \ref{sec:xray}). c) Polarized radio corona density structure (Section \ref{sec:radio}). d) Predicted radio emission from the ECM instability. e) Light curve of X-ray variability and radio variability. The radio intensity peaks when the dipole axis is in the plane of the sky for the observer and the maximum volume of the X-ray emitting corona is eclipsed by the star.}
\label{fig:dipole}
\end{figure*}

Figure \ref{fig:dipole}a shows a simulated magnetic map of a simple, inclined dipole. For this model the peak magnetic field strength of the dipole is set to $B_0=1,000$ G and the inclination of the dipole axis, $\beta=40^\circ$.
Over plotted are the results of applying the PFSS model (Section \ref{sec:pfss}) with the closed field lines shown in red and the open, wind bearing loops is shown in blue. Figure \ref{fig:dipole}b shows the X-ray emitting corona for the inclined dipole (Section \ref{sec:xray}). In this simulation we have assumed a coronal temperature of $T_{\rm cor} = 5\times10^6$ K, which is typical for rapidly rotating stars \citep{johnstone2015}. Figure \ref{fig:dipole}c shows the regions of the corona that satisfy the conditions for ECM (Section \ref{sec:radio}), where we have color coded the emission based on the polarity of the radio photons, which is determined by the sign of the local magnetic field, with red being positive and blue being negative. Finally, Figure \ref{fig:dipole}d shows the radio spectrum for the inclined dipole. The two bright peaks in the spectrum that occur at longitude $90^\circ$ and $270^\circ$ correspond to times when the inclined dipole is in the plane of the sky, since the ECM emission is emitted at $90^\circ$ to the magnetic field line. Since the magnetic field is a \replaced{pure dipole}{dipole with a source surface, the field strength as a function of distance from the stellar surface can be expressed as,
\begin{equation}
    B(r) = \frac{2M\cos\theta}{r^3}\left(\frac{r^3+2r_{\rm ss}^3}{r_\star^3+2r_{\rm ss}^3}\right),
    % B(r) \approx B_0\left(\frac{r_\star}{r}\right)^3.
\end{equation}
where $M=B_r(r=r_\star, \theta=0)r_\star^3/2$ is the dipole moment for a purely dipolar field \citep{jardine2002}.}
Since the frequency of the ECM emission is directly related to the magnetic field strength, we can determine the maximum frequency of the radio emission, $\nu_{\rm max} = 2.8B_0\simeq2.8$ GHz.

Since we have computed the X-ray and radio coronal densities, we can compare their observable light curves. Figure \ref{fig:dipole}e shows the X-ray variability and also the radio variability (at 1.2 GHz) as a function of stellar longitude. The light curves clearly show that the X-ray variability is anti-phased with the radio emission, with a Pearson correlation coefficient of $\rho=-0.87$. This anti-correlation occurs because of the field geometry: the radio intensity peaks when the dipole axis is in the plane of the sky for the observer and the maximum volume of the X-ray emitting corona is eclipsed by the star. The longitudes of the the peaks in the radio light curve (Figure \ref{fig:dipole}e) can be shown to be:
\begin{equation}
    \phi = \arccos \left(\frac{-\tan i}{\tan(\alpha + \beta)}\right),
\end{equation}
where $i$ is the stellar inclination, $\beta$ is the angle between the magnetic and rotation axes and $\alpha$ is the angle of the ``auroral oval'', which for a dipole field is given by $\sin^2 \alpha = r_\star/r_{\rm ss}$ .

\subsection{V374 Peg}
\begin{figure*}
\centering
    \includegraphics[width=1\textwidth]{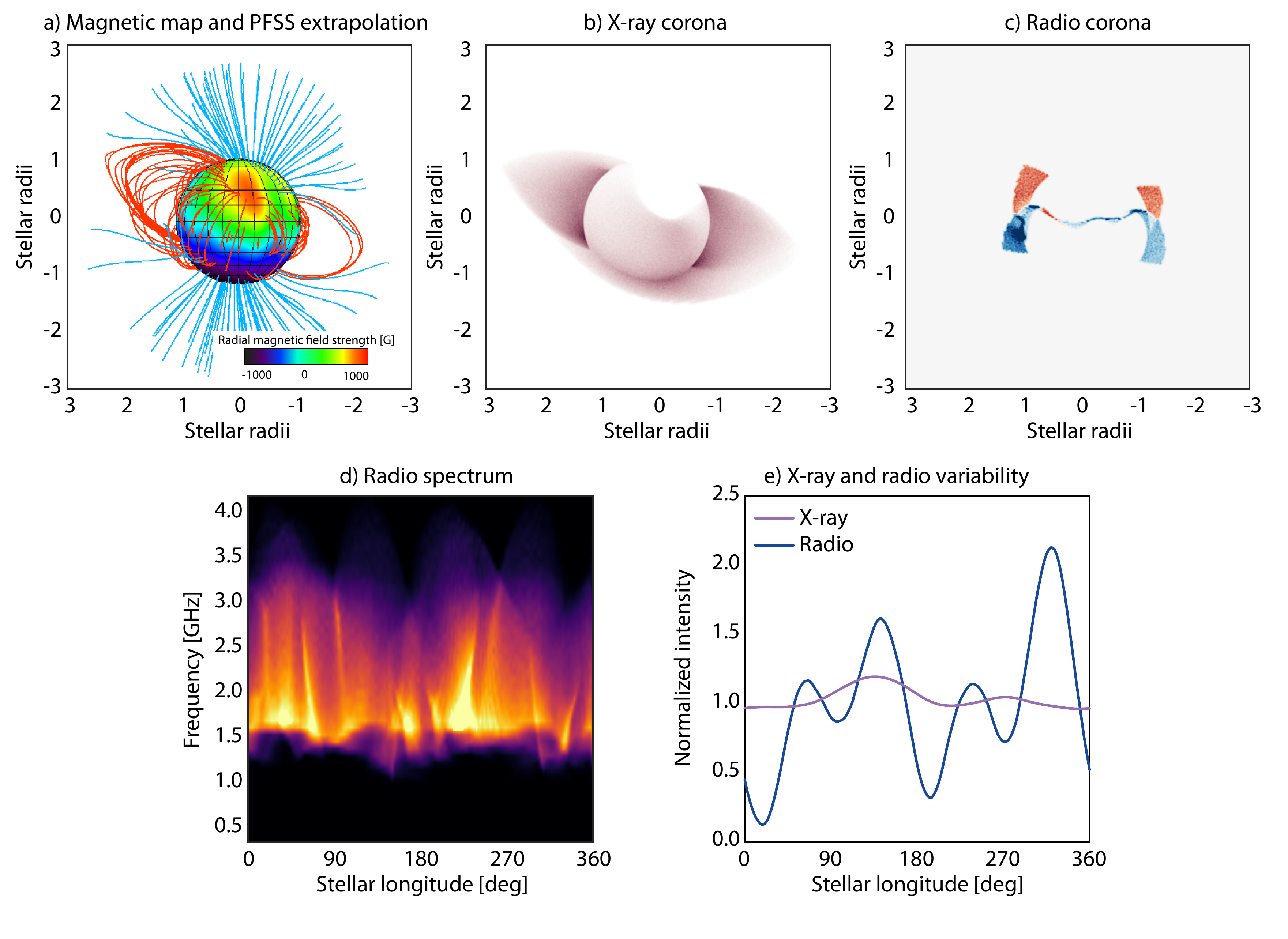}
    \caption{Same as Figure \ref{fig:dipole} but for V374 Peg using the observed ZDI map \citep{donati2006,morin2008v374} as input to the model. Since the observed magnetic field of V374 Peg is more complex than a simple dipole the simulated X-ray and radio coronae show more structure.}
\label{fig:v374}
\end{figure*}
We are interested in determining the variability and frequency of radio emission that originates through the ECM instability for stars using their observed magnetic maps. Figure \ref{fig:v374}a shows the ZDI map of V374 Peg as reconstructed by \citet{morin2008} and the PFSS extrapolation. It is worthy of note that the inclination of the star is such that co-latitudes $\gtrsim120^\circ$ are not visible as the star rotates and therefore the magnetic field cannot be reliably reconstructed in that part of the stellar disk.

Before we can compute the X-ray corona for V374 Peg we must specify the temperature of the corona, $T_{\rm cor}$, and the value for $\kappa$, in the expression for the pressure at the base of each magnetic field line $\left(p_0=\kappa B_0^2\right)$. Both the coronal temperature and base pressure will alter the resultant X-ray luminosity predicted by our model. We can therefore use observations of the X-ray luminosity to better constrain these values. X-ray observations from {\sc Rosat} of V374 Peg measured the X-ray luminosity to be $\log L_X=28.44$ erg s$^{-1}$ \citep{hunsch1999}. To set the temperature of the corona we use the relations derived by \citet{johnstone2015}, where they show,
\begin{equation}
    T_{\rm cor} = 0.11\times10^6F_X^{0.26},
    \label{eqn:tcor}
\end{equation}
where $T_{\rm cor}$ is the coronal temperature in MK and $F_X$ is the X-ray flux in ${\rm erg}\,{\rm s}^{-1}\,{\rm cm}^{-2}$. For V374 Peg using the values from \citep{hunsch1999} we estimate a coronal temperature for V374 Peg of $T_{\rm cor}\simeq6\times10^{6}$ K. Using this value of $T_{\rm cor}$ we then varied the value of the scaling parameter, $\kappa$ to find the best fit to the observed $L_X$. Figure \ref{fig:v374}b shows the X-ray corona when our best fit value of $\kappa$ is adopted.

Figure \ref{fig:v374}c shows the results of applying the model developed in Section \ref{sec:radio} to determine the locations in the corona of V374 Peg that satisfy the conditions for radio emission through the ECM instability (Equation \ref{eqn:condition}). The emission is color coded by the corresponding polarization of the emission with red being positive and blue being negative.

While the magnetic field topology of V374 Peg is predominantly dipolar, the ZDI map does show more structure than the simple dipole shown in Figure \ref{fig:dipole}. This more complex field structure manifests in a more structured X-ray and radio corona. This can be seen most clearly in the radio spectrum (Figure \ref{fig:v374}d) and the X-ray and radio light curves (Figure \ref{fig:v374}e). As with the simple dipole field, the X-ray and ECM light curves are anti-phased; however, due to the increased complexity in the magnetic field, the anti-correlation is not as strong, with a Pearson correlation coefficient of $\rho=-0.44$.

\section{Modeling the radio observations of V374 Peg}\label{sec:v374_obs}
\begin{figure*}
    \centering
    \includegraphics[width=1\textwidth]{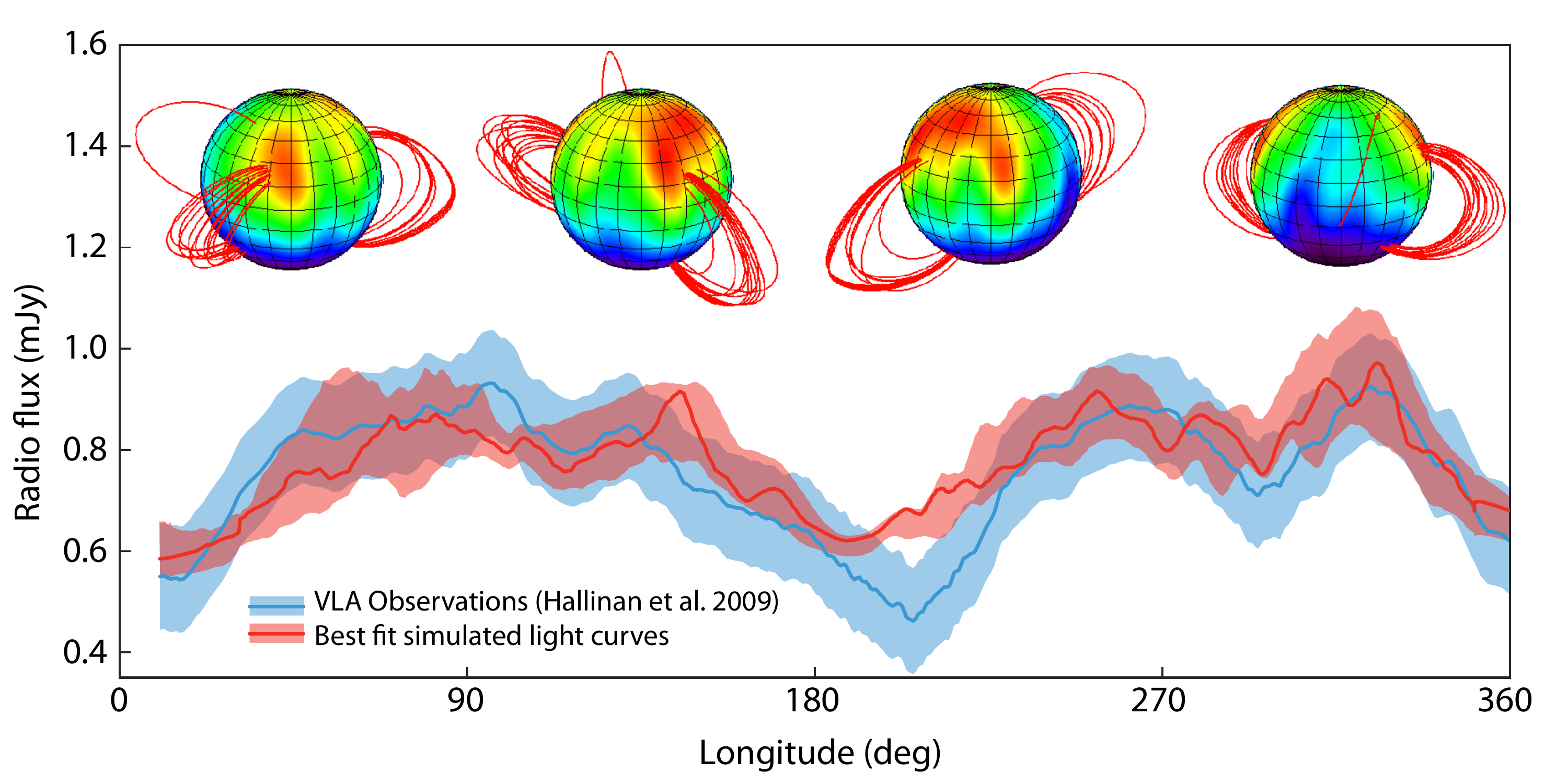}
    \caption{\replaced{Radio observations taken over three nights}{Averaged radio light curve from three nights of observations} of V374 Peg phased to the stellar rotation period ($p_{\rm rot}=0.44$ d) from the Very Large Array (blue) with error (shaded blue). \added{The error is a combination of statistical noise in the individual measurements, and the systematic variation from averaging the three nights of observation.} Here we have removed the pulsed radio bursts and plot only the rotationally modulated, but smoothly varying component of emission. Also shown are \added{all the simulated light curves that provide an equally good fit within error (shaded red) and their average (red) of V374 Peg using our model for ECM emission and the ZDI map (red). Also shown is the PFSS extrapolation for the emitting field lines that are present in over 90\% of our best fit light curves. These field lines determine the overall phasing of the broad modulation of the radio light curve.}}
   \label{fig:radio_fit}
\end{figure*}
V374 Peg was observed for 12 hours on three successive nights at the Very Large Array (VLA) on 2007 January 19, 20, 21, spanning three rotations of the star \citep{hallinan2009}. The observations were obtained using the X band configuration of the VLA, which spans $\nu=4 - 8$ GHz and is therefore sensitive to a magnetic field of $B\sim2,800 - 4,300$ G. A summary of the radio observations, phased to the rotation period of V374 Peg are shown in Figure \ref{fig:radio_fit}. In this light curve we have removed the pulsed radio emission and have only plotted the rotationally modulated but smoothly varying component of the radio emission, which we are attempting to model here.

These observations were taken within just a few months of the ZDI observations. If the origin of this radio emission was ECM then our model should be able to reproduce the radio light curve. The radio spectrum shown in Figure \ref{fig:v374} is the result of assuming every field line that satisfies the conditions for ECM does indeed emit radio photons. In reality it is not necessarily the case that every field line is constantly emitting radio photons. To fit to the radio observations we carried out a Monte-Carlo simulation, allowing a random subset of the field lines capable of emitting ECM photons to do so. In total we ran 100,000 simulations in an effort to determine the best configuration of emitting field lines to match the VLA observations of V374 Peg.

% In our Monte-Carlo simulations we also allowed the inclination of the star to vary from the observed value of $70^\circ$ by $\pm10^\circ$. After 100,000 simulations we found that the best fit to the observations was achieved for an inclination of $65^\circ$.
\replaced{The best fit light curve is shown as the red line in Figure \ref{fig:radio_fit}. The light curve produced by the model reproduces both the observed radio modulation and amplitude, providing evidence that the radio emission observed from V374 Peg could indeed be due to the ECM instability.}{We find that there is not a single configuration of emitting field lines that fit the observations; rather, we find many configurations that are capable of providing an equally good fit to the data. All simulations that show an equally good fit (within error) to the observations are shown as the shaded red curve in Figure \ref{fig:radio_fit}, with the average light curve shown as the solid red line. To investigate which field lines are contributing to the phasing of the broad modulation and which to the amplitude of the radio light curve we isolated those field lines that are common to over 90\% of our best fit simulations. These field lines are shown on the PFSS extrapolations in  Figure \ref{fig:radio_fit}. We find that the common field lines are grouped into two distinct longitude regions separated by $\sim180^\circ$. It is these field lines that determine the phasing of the broad modulation in the radio observations. The number of field lines that are lit, coupled with the choice of other field lines that are not shown in these plots then determines the amplitude of the light curve. }

\added{To further test whether the configuration of the magnetic field determines the phasing and modulation of the radio light curve shown in Figure \ref{fig:radio_fit} we ran multiple simulations where we phase-shifted the observations and then found a new best fit. We found that phase-shifting the observations resulted in very poor fits to the data, and for large shifts ($> 60^\circ$) we were unable to find a fit at all. These additional tests suggest that the magnetic field configuration is indeed responsible for the modulation observed in the radio light curve. We also tested the role refraction may play in altering the shape of the simulated radio light curve by varying the opening angle of the emission cone from $90^\circ$. We found an equally good fit for opening angles $>60^\circ$ suggesting refraction is unlikely to be playing a critical role.}

There are some caveats to our model that are worthy of note. Firstly, the magnetic map and radio observations was not obtained simultaneously, but were obtained within a few months. This is not so critical for modeling the rotationally modulated background emission since multi-epoch observations of V374 Peg have shown the magnetic field to be stable over this time-scale \citep{donati2006,morin2008v374}. However, the lack of simultaneity and the assumption in the ZDI reconstruction process that the magnetic field remains static does hinder our ability to model the pulsed radio emission. Secondly, the radio observations were observed in the X band, which covers $\nu=4 - 8$ GHz ($B\sim1,400 - 2,800$ G); however, in the ZDI map, the maximum magnetic field strength is $B\sim1,660$ G, which means we will only simulate ECM photons at a maximum frequency of $\nu\simeq4.6$ GHz. Underestimating the magnetic flux in a ZDI map is a well known issue and is a consequence of the reconstruction technique being less sensitive to small, strong regions of magnetic field (e.g., \citealt{lang2014}).

\added{While our model only allows for ECM emission from parts of the corona where the plasma frequency is less than the cyclotron frequency (Equation \ref{eqn:ecm1}), it is currently unable to account for the bursty nature of the emission and assumes steady state emission from all the ECM-capable zones. In all our simulations we find that a group of field lines are responsible for the phased modulation in the light curve (PFSS extrapolations in Figure \ref{fig:radio_fit}). Since there are multiple field lines in these regions, our model is unable to differentiate between a single field line emitting constant levels of ECM emission, or a number of the field lines emitting in a bursty fashion.   }

\section{Summary and Discussion}
% \begin{enumerate}
%     \item Discuss how simultaneous ZDI and radio observations could be used to model radio emission.
% \end{enumerate}
We have developed the first model for predicting the frequency, amplitude, and rotational variability of radio emission arising through the  electron cyclotron maser instability using realistic magnetic maps of low mass stars obtained through Zeeman Doppler Imaging. For stars that have a measurement of the X-ray luminosity our model is capable of predicting the expected frequency and rotational variability of the ECM emission.

We have benchmarked our model using ZDI observations of the bright, rapidly rotating, fully convective, low mass star V374 Peg. This star not only has magnetic maps but was also observed nearly simultaneously in the radio using the VLA. Our model successfully reproduces the amplitude and variability of the observed radio light curve, providing further evidence that the radio emission from this star could be due to the ECM instability.

We have only considered radio emission arising through the ECM instability and not through the gyrosynchrotron emission process. \deleted{To estimate the validity of this approach} We use the G\"udel-Benz relation \added{to estimate the magnitude} of the radio flux from gyrosynchrotron emission alone. The  G\"udel-Benz relation is an empirical correlation between the gyrosynchrotron radio emission and the X-ray luminosity for a wide variety of astronomical sources including cool stars, solar flares, active galactic nuclei, and galactic black holes \citep{gudel1993b,gudel1993a}. The relation can be expressed as,
\begin{equation}
    L_{X} \approx L_{\nu,R}\times10^{15.5},
\label{eqn:gb}
\end{equation}
where, $L_X$ is the observed X-ray luminosity of the source, and $L_{\nu,R}$ is the radio luminosity from gyrosynchrotron emission alone. Using the  G\"udel-Benz relation and the observed X-ray luminosity of V374 Peg ($L_X=10^{28.44}$; \citealt{hunsch1999}), V374 Peg's radio luminosity from gyrosynchrotron emission alone should be $L_{\nu,R}=10^{12.94}$. Using the distance to V374 Peg ($d=8.93$ pc; \citealt{vanleeuwen2007}), this luminosity corresponds to a radio flux of $F_X\sim 0.08$ mJy. From the VLA observations (Figure \ref{fig:radio_fit}) the observed radio flux is at least one order of magnitude higher than this value, suggesting that gyrosynchrotron emission is a negligible contribution to the total radio flux from V374 Peg. \added{It is worthy of note that there is uncertainty in the G\"udel-Benz relation, particularly for low mass stars and ultracool dwarfs which appear to lie above this relation.}

\added{Simultaneous VLA and \textit{Chandra} observations of the Orion Nebula Cluster by \citet{forbrich2017} enabled these authors to search for correlations between extreme radio and X-ray variability from young stellar objects. They found 13 radio sources, all of which also exhibited X-ray variability. Multi-epoch radio, optical (including H$\alpha$), UV (\textit{Swift}), and X-ray (\textit{Chandra}) observations of the UCD binary NLTT 33370 AB by \citet{williams2015} found periodic modulation in the radio and optical and plausible modulation in H$\alpha$ and the UV.} Comparing simultaneous X-ray light curves with radio observations may help assess the relative contributions of radio emission through ECM and gyrosynchrotron processes. If the dominant source of radio emissions is through the ECM instability as modeled here, the radio and X-ray light curves should be anti-phased; however, if the dominant emission process is gyrosynchrotron emission then the light curves should be phased.

\deleted{From our simulations, we find a strong antisymmetry between the X-ray and ECM light curves for a simple dipolar magnetic field configuration. This antisymmetry is less strong for the more complex magnetic field of V374 Peg, suggesting that future observations of both radio and X-ray light curves could potentially be used to quantify the complexity of a stellar magnetic fields independently of ZDI observations.}

In the future, this model will be used to predict the expected radio emission from the ECM instability for all low mass stars with a magnetic map and an X-ray luminosity measurement. These predictions will be useful for determining the expected frequencies at which ECM emission is likely to be observed, and will help guide future observations with the Karl G. Jansky Very Large Array. In the search for radio emission from exoplanets, our method could also potentially be used to model the stellar component to help disentangle radio signals from an orbiting exoplanet.

\section*{Acknowledgments}
We would like to thank the anonymous referee for their helpful comments and suggestions. MMJ acknowledges support from STFC grant ST/M001296/1. \added{This research has made use of the SIMBAD database, operated at CDS, Strasbourg, France. This research has made use of NASA's Astrophysics Data System. This research made use of Astropy, a community-developed core Python package for Astronomy \citep{2013A&A...558A..33A}}

% \listofchanges
% \newpage



\bibliography{refs}



\end{document}